# Towards a generalized accessibility measure for transportation equity and efficiency

Rajat Verma, Mithun Debnath, Shagun Mittal, Satish V. Ukkusuri

Lyles School of Civil Engineering, Purdue University, West Lafayette, IN 47907

## Abstract

Locational measures of accessibility are widely used in urban and transportation planning to understand the impact of the transportation system on influencing people's access to places. However, there is a considerable lack of measurement standards and publicly available data. We propose a generalized measure of locational accessibility that has a comprehensible form for transportation planning analysis. This metric combines the cumulative opportunities approach with gravity-based measures and is capable of catering to multiple trip purposes, travel modes, cost thresholds, and scales of analysis. Using data from multiple publicly available datasets, this metric is computed by trip purpose and travel time threshold for all block groups in the United States, and the data is made publicly accessible. Further, case studies of three large metropolitan areas reveal substantial inefficiencies in transportation infrastructure, with the most inefficiency observed in sprawling and non-core urban areas, especially for bicycling. Subsequently, it is shown that targeted investment in facilities can contribute to a more equitable distribution of accessibility to essential shopping and service facilities. By assigning greater weights to socioeconomically disadvantaged neighborhoods, the proposed metric formally incorporates equity considerations into transportation planning, contributing to a more equitable distribution of accessibility to essential services and facilities.

# 1   Introduction

Accessibility refers to the ease of reaching desired destinations, services, or activities within a given environment [1]. It is a fundamental aspect that influences social inclusion, happiness, and overall quality of life. In recent years, the recognition of the negligence of accessibility to low-income and historically underserved communities has given rise to a shift towards prioritizing accessibility over mere mobility [2]. This shift recognizes that a well-connected and efficient transportation system does not guarantee that everyone can access essential opportunities like jobs, hospitals, schools, and healthy food stores [3]–[5]. Factors like the spatial distribution of resources, the affordability of transportation options, and the needs of different demographic groups all play a role in determining accessibility [6], which consequently connects it with the issue of social equity [7]. The current state of practice of accessibility measurement does not inherently consider the relationship between the built environment, which includes the spatial distribution of opportunities, land use, and the transportation system, with equity issues of socio-economically disadvantaged communities [8], [9]. A significant challenge for urban planners is a lack of standardization of accessibility measurement [1], [6] to improve decision-making for the investment of infrastructure towards higher accessibility for all. This is because accessibility studies are often conducted for particular cities or metropolitan areas using specific formulations of accessibility, which results in disparate and sometimes incompatible assessments across studies [6], [10]–[12]. We identify two major components of this problem – (i) differences in approach to measurement and (ii) limited data availability.

First, issues in measurement stem primarily from differences in the understanding of accessibility as a function of the built environment and travel behavior.  For example, the traditional contour or cumulative opportunities measure assumes that places are accessible as long as they fall within a catchment defined by a specific threshold of travel impedance, usually measured by distance, time, or monetary cost [10]. This assumption does not reflect travel behavior adequately since it weights faraway places the same as nearby places and thus violates the basis of spatial interaction models [13]. This has consequences as urban planners may deprioritize investment in rural and suburban regions since this approach overestimates accessibility in the periphery of metropolitan areas [14]. Gravity-based measures partially solve this problem by considering travel impedance functions, but their continuous nature does not provide actionable insights like identifying transportation infrastructure efficiency for policies like 15 or 30-minute cities [12]. This is aggravated by the observation that impedance changes significantly by travel mode and trip purpose [15].

Issues in measurement also stem from differences in prioritization. Traditional accessibility measurement is aimed at maximizing utility in terms of reachability to places [16]. However, core urban areas dominate reachability maximization for most people, without consideration for necessities by all people [17]. A lack of explicit integration of social equity concerns in this utility function is another important cause of concern for accessibility measurement for urban planners [17].

Second, there is a lack of general purpose large-scale publicly available data for accessibility analysis. Existing comprehensive datasets of accessibility are limited in important ways. WalkScore, for example, is a popular company that provides accessibility scores at the street address-level by walking, bicycling, and public transit [18]. However, its scoring methodology is not transparent, it only caters to accessibility to arbitrarily defined categories of points of interest (POIs), and does not include accessibility to jobs, an important component widely discussed in the research literature [19]–[22]. On the other hand, the large-scale granular accessibility dataset prepared by Owen et al. [23] pertains to only jobs and does not include accessibility to important POIs such as grocery stores, hospitals, and schools, nor is their code openly accessible for peer review and improvement. This unavailability of data and reproducible methodological tools across different regions limits comprehensive and comparative analyses of accessibility.

To address these two main challenges, we propose a generalized accessibility metric that combines the notions of contour and gravity measures by introducing thresholds to empirically fitted impedance functions. This metric is computed for multiple travel modes, trip purposes, and travel time thresholds. We make public a dataset that includes accessibility values computed for each census block group of the continental United States (US) and the source code associated with its computation. We further contribute to the literature on accessibility analysis by analyzing the relationship between accessibility and the built environment concerning two concepts – transportation infrastructure efficiency measurement and equity-based investment prioritization. The proposed generalized accessibility metric and open dataset will assist research and practice on standardized measurement. It will also enable a large-scale assessment of the transportation system's role in providing equitable access to opportunities.

The rest of the paper is organized as follows. Section 2 provides a review of the state of practice of accessibility measurement. Section 3 shows the methodological framework of this study, including the definition of the proposed metric, a brief description of input datasets, and assessment of the parameters such as trip purpose, mode, and threshold. Section 4 discusses the relationship of this accessibility metric with the built environment considering the notions of efficiency and equity. Section 5 summarizes the findings and provides insights into the application of this accessibility metric in practice.

## 2  State of practice

In urban and transportation planning, mobility and accessibility are often considered mutually exclusive paradigms, particularly for the task of benefits assessment [9], [24]. Mobility studies are primarily concerned with the physical movement, speed, and efficiency of transportation systems, with a focus on performance metrics like speed, delay, and level of service [25]. Accessibility, on the other hand, is the ease and extent to which opportunities (such as jobs, education, healthcare, etc.) can be visited within some cost constraints [26]. The research literature on accessibility measurement is large and diverse. Broadly speaking, current measures of accessibility can be classified as either subjective/objective or normative/positive [10]. Subjective measures reflect people's opinions or perceptions of their environment, while objective measures reflect the actual circumstances in which people live and work [27], [28]. Similarly, normative measures aim to quantify the potential of accessibility that people *ought to* have as opposed to positive measures which quantify people's *observed* access to opportunities, usually measured by realized mobility and employment [10]. Subjective measures are usually derived from travel surveys and are thus limited by the limitations of scalability, lack of longitudinal data, and recall bias [8]. For large-scale analysis, objective measures are more advantageous and are more commonly used [10].

Among objective measures, there are five broad categories – contour, gravity-based, utility-based, person-based, and space-time measures [12], [26]. The first two categories are also sometimes called locational or spatial measures and are widely used in literature due to their ease of computation and scalability. The latter three categories of measures are not easily scalable to large areas at high fidelity of spatial units. For example, utility-based measures estimate the value of opportunities based on the assumption that the users seek to maximize the utility of their behavioral choices, such as for modes and destination types [26]. Their discrete optimization formulation greatly limits their scalability [10]. Person or individual-based measures consider the distributions of travelers' travel capabilities and preferences, activity schedules, and space-time constraints in computing regional accessibility [29], [30], making them sophisticated but also difficult to collect representative data for large regions. Space-time measures, such as space-time prisms, limit the opportunities reachable within not only spatial but also temporal constraints, such as minimum time spent at a place [31]–[33]. These are also often limited to disparate cities and small regions due to a lack of availability of temporal data like operating hours of stores or public transit schedules [31]. For these reasons of scalability, we do not consider these classes of measures in our analysis.

Contour and gravity measures are similar in their measurement in that they only require the spatial distribution of travel demand generators (usually residences) and attractors (also called opportunities), such as workplaces, shops, and facilities. They both involve computing the total number (or weight) of opportunities reachable from a given origin place by considering the cost or 'impedance' between it and the candidate destinations. Thus, the general formulation of accessibility, $a_i$, of a place $i$ is:

$$a_i = \sum_j o_j \cdot f(x_{ij}) \tag{1}$$

Here, $o_j$ is the total number or weight of opportunities in the destination place $j$. The cost, $x_{ij}$, refers to the 'disutility of travel' between places $i$ and $j$. It is usually measured in terms of distance, travel time, fare, energy consumption or discomfort [15], [34], or a combination thereof (e.g., [35]). $f$ is called the weight or impedance function that maps the change in travel disutility to the change in travel likelihood.

The contour measure, also referred to as reachability or cumulative opportunities, simply considers the total number of opportunities reachable within the cost constraints [21], [26]. Its impedance is therefore an indicator function, $f(x; \tau) = I_{\{x \leq \tau\}}$, of interzonal cost $x$ and cost threshold $\tau$.

Gravity measures, on the other hand, do not consider a cost threshold per se but assume that the likelihood of travel decreases asymptotically with increasing cost. This inverse proportionality is represented in the literature by 'decay impedance functions' that are often approximated by fitting on the probability distribution of traveling traditionally based on travel surveys [15], [36] or passively determined mobility such as using mobile phone location data [37]. These functions often lie in the family of exponential functions, with notable members including the negative exponential ($f(x) = e^{\theta x}$) [15], [19], power exponential ($e^{ax^b}$) [38], and power function ($x^\theta \equiv e^{\theta \ln x}$) [5].

While gravity measures are more realistic than contour measures since they indicate the number of opportunities 'reasonably' accessible from a place, they cannot be conveniently interpreted as the number of opportunities 'reasonably reachable within cost constraints'. This notion is particularly important in cases where planners need to enforce a design criterion such as a 30-minute target for specific facilities like schools or crèches [10]. In this study, we combine the contour measure approach with the gravity approach by introducing a threshold on decay functions. This effectively turns our metric into a 'catchment' measure where the destination zones in the gravity measure are bounded by the cost threshold considered. The concept of catchment is widely used in practice such as in the 'floating catchment area' family of metrics that include competition effects to workplaces and medical facilities where it is assumed that the destination's attractiveness decreases if other people (competitors) can access it [39]–[42].

# 3 Data and methods

The framework used to develop the proposed accessibility metric and its analysis with the built environment is shown in Figure 1. The metric is defined for multiple kinds of opportunities, including jobs and POIs, travel modes, and travel time thresholds. It can be aggregated at a regional level by weighting zonal accessibility with the zonal populations.

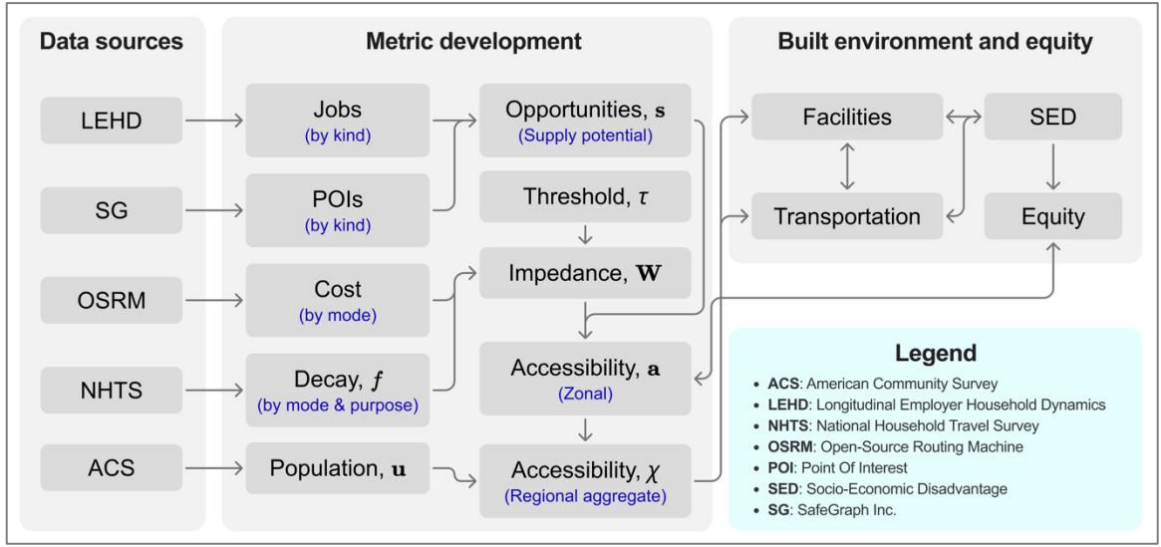

Figure 1: An overview of the accessibility computation and component analysis performed in this study.

## 3.1 Metric definition

We formulate the following general measure of accessibility based on prevalent research literature:

$$a_i^{k,m,\tau} = \sum_{j \in \mathcal{R}} w_{i,j}^{k,m,\tau} o_j^k \qquad (2)$$

Here, the accessibility $a_i$ of a zone or neighborhood $i$ refers to the total number of opportunities of trip purpose or destination kind $k$ accessible from zone $i$ by travel mode $m$ within a cost threshold of $\tau$. $o_j^k$ denotes the supply potential of the destination zone $j$ among all zones of the region $\mathcal{R}$, given by the total number of opportunities of kind $k$ in zone $j$. $w_{ij}$ denotes the impedance weight between the zones $i$ and $j$. This impedance weight is given as a thresholded decay function of the cost $c_{ij}^m$ between $i$ and $j$ that may depend on the mode $m$. It is given by:

$$w_{i,j}^{k,m,\tau}(c_{i,j}^m) = \begin{cases} f^{k,m}(c_{i,j}^m) & c_{i,j}^m \leq \tau \\ 0 & c_{i,j}^m > \tau \end{cases} \forall i,j \in \mathcal{R} \qquad (3)$$

The metric in Eq. (2) represents the total number of opportunities accessible from origin zones. These opportunities can only be realized by people (presumably residents) of that zone. To make sure this metric adequately represents the overall accessibility, it would be reasonable to weight the values of this metric by the total number of people or employees of the zones when aggregating over larger geographies [23]. For a subregion, $\mathcal{S}$, say a county of MSA, of the region $\mathcal{R}$, the aggregate accessibility, $\chi_\mathcal{S}$, can be computed as:

$$\chi_\mathcal{S}^{k,m,\tau} = \frac{\sum_{i \in \mathcal{S}} a_i^{k,m,\tau} n_i}{\sum_{i \in \mathcal{S}} n_i} \quad (4)$$

Here, $n_i$ is the total population or number of workers of zone $i$. By defining a population weight vector $\mathbf{p}_\mathcal{S} = \left\{ p_i = \frac{n_i}{\sum_{j \in \mathcal{S}} n_j} \right\}_{i \in \mathcal{S}}$ for region $\mathcal{S}$, the equations (2) through (4) can be combined succinctly as:

$$\chi_\mathcal{S}^{k,m,\tau} = \mathbf{p}_\mathcal{S}^T \mathbf{a}_{\mathcal{S},k,m,\tau} = \mathbf{p}_\mathcal{S}^T \mathbf{W}_{\mathcal{S},k,m,\tau} \mathbf{o}_k \quad (5)$$

Here, $\mathbf{a}_\mathcal{S}$ is the vector of accessibility of all zones in the target region $\mathcal{S}$, $\mathbf{W}$ is the impedance weights matrix of size $|\mathcal{S}| \times |\mathcal{R}|$, and $\mathbf{o}$ is the opportunities vector for the larger region $\mathcal{R}$. Note that the accessibility of a region depends on all its surrounding zones, which is a theoretically unbounded set. For practical reasons of computation, the environment set $\mathcal{R}$ is limited to a province, state, or MSA, since the weights matrix with a maximum size is $|\mathcal{R}|^2$ can be computationally expensive.

In this paper, we discuss one specific form of accessibility that uses only travel time as the cost measure and the power exponential impedance function, $f(x) = \exp(-\alpha x^\beta)$. Using this information and combining Eq. (2) and (3), we get:

$$a_i^{k,m,\tau} = \sum_{j \in \mathcal{R}:\, t_{i,j}^m \leq \tau} o_j^k \cdot \exp\left(-\alpha_{k,m} \left(t_{i,j}^m\right)^{\beta_{k,m}}\right) \quad (6)$$

The parameters $\alpha$ and $\beta$ are fitted empirically for different modes and trip purposes. The values of opportunities count, $o$, and interzonal travel time, $t$, are obtained from data sources explained in the next section.

## 3.2 Data description

For the accessibility dataset prepared in this study, we use the following publicly available datasets as briefly described below, with further details provided in Suppl. Section 1.

**Impedance functions**: Impedance decay functions for travel time of the power exponential form are fitted using the National Household Travel Survey (NHTS) 2017 trips dataset for the US. The data processing details and the fitted parameters computed for five trip purposes for driving, walking, and bicycling are provided in Suppl. Section 1.2.

**Jobs**: Total job counts by census block group (BG) are obtained from the Longitudinal Employer Household Dynamics dataset. In addition to total jobs, counts of jobs of high and low earnings are also collected (see Suppl. Section 1.3).

**Points of interest** (POIs): The POI dataset is obtained from SafeGraph Inc. as part of its public release of data during 2020–21 to aid research related to COVID-19. Its US dataset contains locations of 5.5 million POIs across 92 categories, which are further categorized into three special classes for this study – 'essential stores', 'primary services', and 'leisure' places. The details are provided in Suppl. Section 1.4.

**Interzonal costs**: The distances and travel times between each pair of zones are computed using the Open-Source Routing Machine project [43]. The processing details and the distributions of the distances, travel times, and speeds are provided in Suppl. Section 1.5.

**Socio-economic disadvantage**: Demographic data are obtained from the American Community Survey (ACS) 5-year estimates dataset for 2021. The description of the obtained ACS fields and their processing is described in Suppl. Section 1.6.

## 3.3 Trip purpose and mode

Travel mode and trip purpose can particularly influence the odds of choosing a destination opportunity which is reflected in its accessibility. For example, most places are much less accessible by walking than driving, primarily due to impedance factors like low speed, inconvenience, and safety concerns [44]. Similarly, the impedance of visiting a place also depends on the purpose of the trip [37]. For instance, the likelihood of travel might be much more sensitive to leisure trips by public transit fare price (a measure of impedance) compared to commute trips [45]. We observe substantial differences in the impedance of travel by mode and purpose in the NHTS travel survey data (see Figure 2a), which prompted us to develop different decay functions by mode and purpose. The fitted parameters for these curves are provided in Suppl. Table 2.

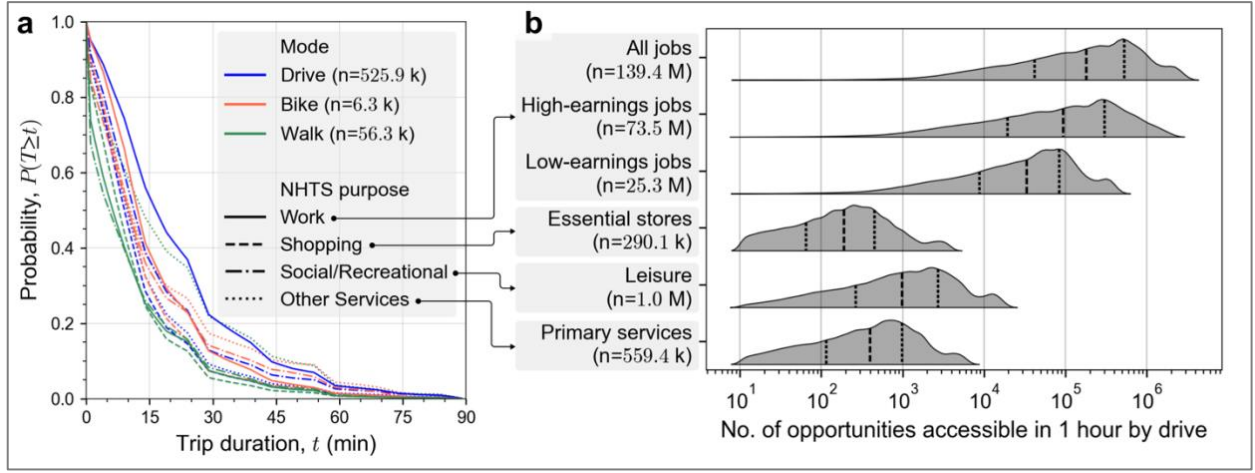

Figure 2: Distributions of trip purpose and the corresponding accessibility. (a) Cumulative probability distribution of trips in the NHTS dataset by mode and purpose (5-minute-smoothened curves). The trip counts by mode are shown in parentheses. (b) Distribution of 60-minute driving accessibility by opportunity kind, across BGs, with arrows showing their mapping to the NHTS trip purposes. The dashed vertical lines denote the quartiles of the distributions and the numbers in parentheses denote the total number of OD pairs in the US dataset.

The distributions of the 60-minute accessibility in the BGs of the US by opportunity kind are shown in Figure 2b. First, it should be noted that comparison of accessibility figures should be focused on differences by zone or region rather than purpose. This is more important in the case of job and POI accessibility, which have very different scales, primarily due to the differences in the underlying datasets. Particularly, access to jobs with high earnings (>$40k per year) is usually much higher than to jobs with low earnings (≤$20k per year), mainly because of more jobs falling in the high-earnings category than the low-earnings category in the LEHD dataset [46]. Also, accessibility to essential stores and services is not only in general lesser than that to leisure places, but their distributions are also skewed more towards the left, indicating a substantial number of BGs not having substantial accessibility to these essential places. This topic is revisited in Section 4.2 where this disparity is shown to be highly correlated with socio-economic disadvantage.

## 3.4 Travel time threshold

Other than the impedance function parameters, the main parameter of the proposed accessibility metric is the cost threshold, $\tau$, particularly the travel time threshold. To test the sensitivity of the results with this threshold, we compute accessibility for five values of this parameter: 15, 30, 45, 60, and 90 min. Referring to the observed distribution of trip duration in the NHTS dataset (Suppl. Figure 1), it appears that the 90-minute threshold may be considered enough to be practically infinite, making the resultant accessibility values equivalent to an unbounded gravity measure.

It should be noted that the cumulative opportunities approach always overestimates the accessibility of a region because it involves assigning equal importance to near and far destinations. For example, at a threshold of 60 minutes, the accessibility by driving in the Chicago MSA is overestimated by as much as 800% when measured using the constant unit weight in the contour measure instead of decaying impedance weights, as seen in Figure 3b. This overestimation is observed to be the highest in remote areas because of the compounding network effect of including already overestimated neighbors in remote areas. This effect is the least in core urban areas but is more spatially smoothened than the base decaying weights accessibility (Figure 3a). Further, this overestimation of accessibility also increases rapidly with increasing travel time threshold (Figure 3c). This is because with increasing threshold, the difference between the observed travel behavior (as in Figure 2a) and the analyst's assumption about accessibility (contour measure with constant, unit impedance weight) increases. This suggests that studies that rely on the contour measure of accessibility should not choose higher thresholds as the deviation from realistic accessibility increases sharply.

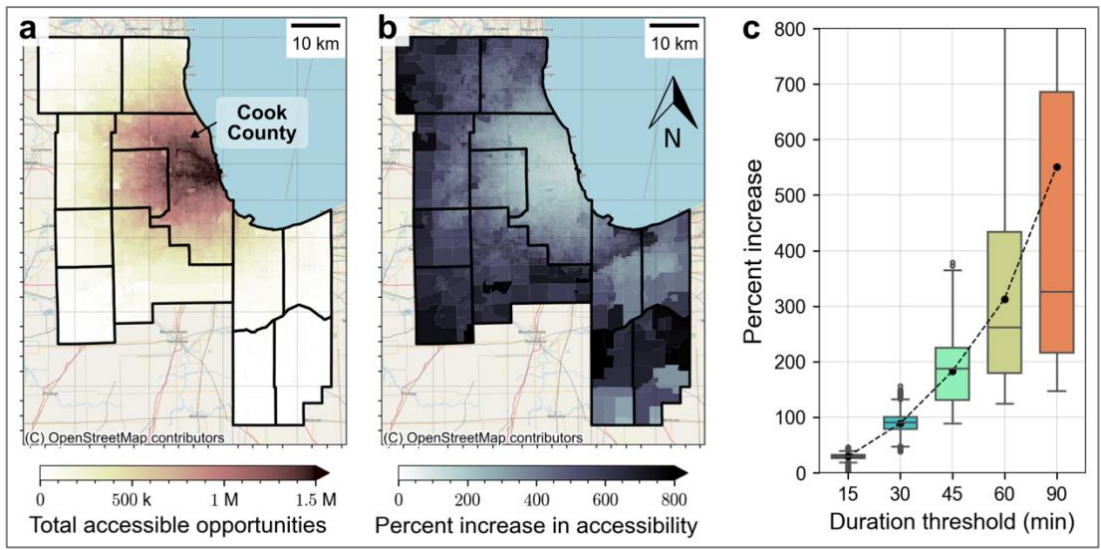

Figure 3: Overestimation of accessibility in the BGs of Chicago MSA by using the contour measure compared to gravity-based decaying impedance. (a) Map of job accessibility using the base decaying impedance function $f(t) =$

$e^{-0.008t^{1.467}}$ within 60 minutes; (b) Percent accessibility overestimation by using the contour measure with $f(t) = 1$; (c) Accessibility overestimation for different duration thresholds. The dashed curve connects the mean values.

## 4 Accessibility, equity, and built environment

### 4.1 Transportation efficiency

Accessibility is directly a function of the transportation system by the inclusion of routed travel times by different travel modes. In this section, we assess the impact of the efficiency of the transportation system of a region on its accessibility while other investments of the built environment (namely the opportunities of interest) are assumed to be constant.

Referring to Eq. (5), the weight matrix **W** is influenced only by the transportation infrastructure of the study region, which in turn depends on its land use distribution. Both the population weights vector, **p**, and the opportunities vector, **o**, depend on land use as well. The role of the transportation infrastructure in accessibility can be better understood by assuming a hypothetical ideal case and introducing the notion of efficiency. Suppose the best case for mobility is a completely flat, unhindered land where travelers can move at up to a given maximum speed that depends on the travel mode. Assuming the same distribution of population and opportunities in the region, the maximum possible accessibility of the region $\mathcal{S}$ can be computed as:

$$\hat{\chi}_{\mathcal{S},k,m,\tau} = \mathbf{p}_{\mathcal{S}}^T \widehat{\mathbf{W}}_{\mathcal{S},k,m,\tau} \mathbf{o}_k \tag{7}$$

The maximal weights matrix is independent of the transportation infrastructure. The minimum travel time between a zone pair $(i,j)$ is given by the Haversine distance between the centroids of the zones divided by a presumed maximum modal speed, $\hat{v}_m$. Using the same impedance function, $f^{k,m}$, as used for computing **W**, the maximal weight for zone pair $(i,j)$ is given by:

$$\widehat{w}_{ij}^{k,m,\tau}(d_{ij}) = \begin{cases} f^{k,m}\left(\frac{d_{ij}}{\hat{v}_m}\right) & d_{ij} \leq \tau \cdot \hat{v}_m \\ 0 & \text{else} \end{cases} \tag{8}$$

We now define 'accessibility efficiency', $\eta$, as the ratio of observed potential accessibility to the maximum possible value when there is no impedance to travel other than the modal speed limit. Theoretically, it represents how well the current transportation infrastructure utilizes the land use distribution that defines the distances between the zones, similar to the work of Dong et al. (2016) [47]. Dividing equations (7) and (8) and simplifying the notation for a selected tuple of $(\mathcal{S}, k, m, \tau)$, we get:

$$\eta := \frac{\chi}{\hat{\chi}} = \frac{\mathbf{p}^T \mathbf{W o}}{\mathbf{p}^T \widehat{\mathbf{W}} \mathbf{o}} \tag{9}$$

This efficiency factor can be used to compare the transportation component of accessibility across different regions (e.g., cities, counties, states), travel modes, and trip purposes. Assuming constant distribution vectors **u** and **s**, an efficiency factor can similarly be computed for each zone $i$ as $\eta_i = a_i/\hat{a}_i$, whose visualization can be useful in inspecting subregional differences in the utilization of transportation infrastructure. We illustrate this concept by comparing three large metropolises of the US – Houston, New York City (NYC), and Chicago. NYC and downtown Chicago are known to be highly compact cities whereas Houston is known for its sprawling infrastructure [48].

The differences in these cities' accessibility efficiency are shown in Figure 4, computed with maximum modal speeds of 60, 4, and 16 mi/h for driving, walking, and bicycling respectively. Panel row (a) shows the zonal efficiency (as a percentage) by driving under 30 minutes. The downtown areas make the most efficient transportation use across all the cities, highlighted by the regions shaded blue. NYC and Chicago have much higher efficiency than Houston, which directly corresponds to compactness and sprawl. In Chicago, the zonal efficiency is also high along the highway corridors of I-90 and I-290, as previously observed in Figure 3. Further, this efficiency is also significantly higher for work-related trips by driving compared to other purposes such as trips to essential services (like schools, hospitals, and grocery stores) (panel row (c)). It may be argued that the transportation infrastructure in these cities is more targeted for commuting by driving than by other modes and for other purposes, especially essential services.

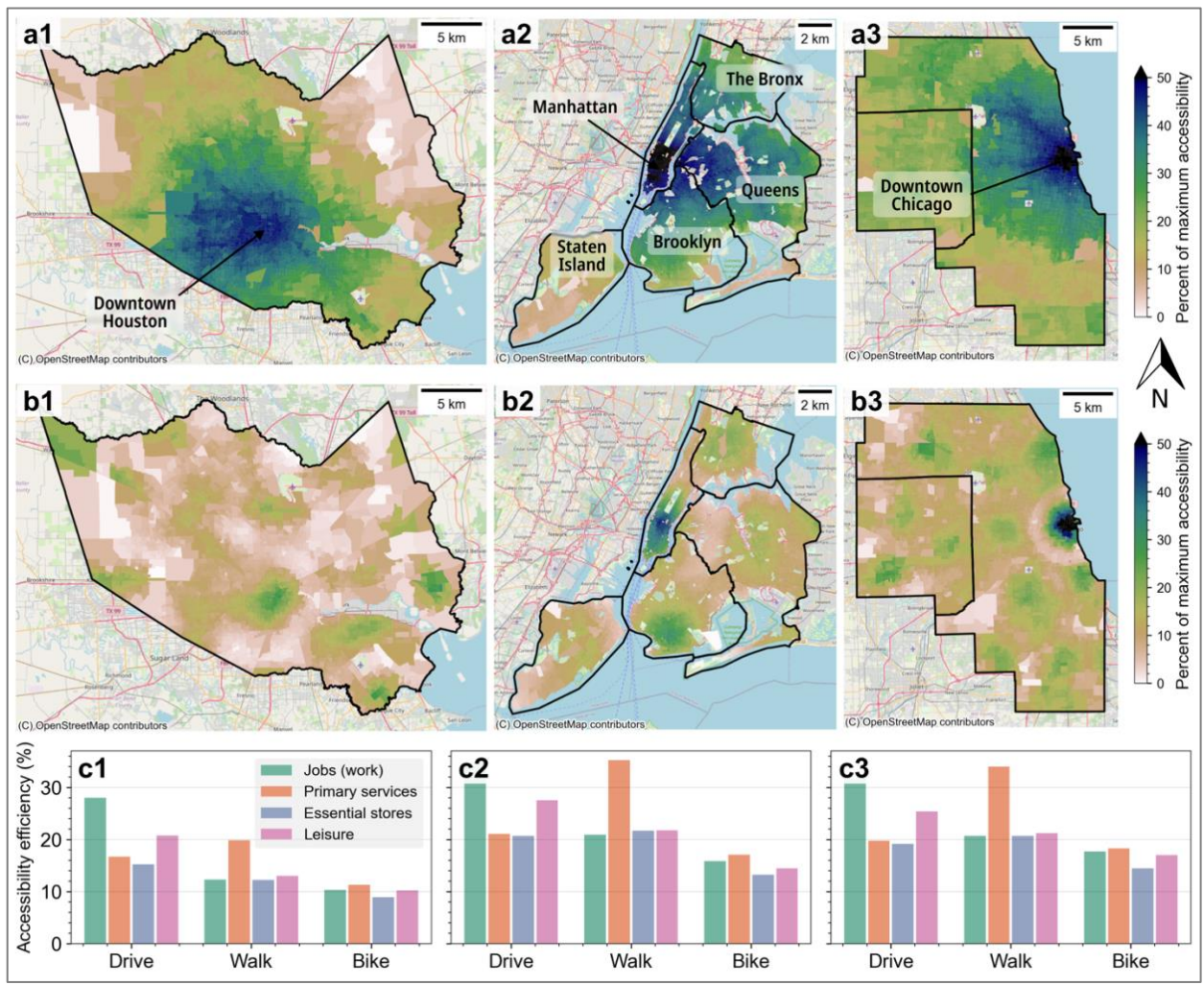

Figure 4: Accessibility efficiency, $\eta$, by (a) driving and (b) bicycling within 30 minutes at the BG level in three urban areas (columns): (1) Harris County of Houston MSA, (2) New York City, NY, and (3) Cook and DuPage counties of Chicago MSA. (c) Aggregate 30-minute accessibility efficiency in these cities by mode and trip purpose.

Panel row (b) shows the spatial distribution of efficiency by bicycling. Houston appears to have the most inefficient bicycle infrastructure, with even the most well-connected regions of downtown Houston utilizing only up to about 25% of their maximum accessibility. NYC fares better in central Manhattan and southern Brooklyn but has patches of inefficiency in northern Brooklyn, southern parts of the Bronx, and most of Staten Island. Chicago has a prominently contoured core of high bike accessibility efficiency downtown, with higher peak efficiency than in New York and Houston. This core declines rapidly in prominence in a ring outside downtown, indicating a sudden degradation of the bicycle infrastructure in this ring. In the next section, we show that there are equity issues surrounding this ring of efficiency decline.

## 4.2 Equitable investment placement

The optimal placement of new businesses and facilities for utility maximization is a key task of urban planning [49], [50]. For transportation planners, a core component of this utility is accessibility [26]. For specific facilities considered essential, such as schools, grocery stores, and hospitals and clinics, careful consideration must be given to historically underserved and overburdened communities, which are often located in socio-economically disadvantaged neighborhoods [51]. The conventional approach to address equity concerns related to business or facility investments involves looking at the impact of investment on local communities' post-location decisions rather than deciding the investment placement beforehand [52]. This oversight often stems from an inadequate understanding of the network effects associated with investments.

Our proposed aggregate accessibility measure can assist this planning decision by understanding the marginal contribution of investment of opportunities in a zone to the overall region's accessibility. Referring to Eq. (5) where the aggregate accessibility of a region $\mathcal{S}$ is expressed as $\chi_\mathcal{S} = \mathbf{p}_\mathcal{S}^T \mathbf{W}_\mathcal{S} \mathbf{o}$, it can be deduced that the overall increase in regional accessibility, resulting from an increase in zonal opportunity counts by $\Delta \mathbf{s}$ is computed as:

$$\Delta \chi_\mathcal{S} = \nabla_\mathbf{o} \chi_\mathcal{S} \circ \Delta \mathbf{o} = (\mathbf{W}_\mathcal{S}^T \mathbf{p}_\mathcal{S})^T \Delta \mathbf{o} \qquad (10)$$

Assuming that a planning authority of the region $\mathcal{S}$ can only intervene within its jurisdiction, it follows that elements of $\Delta \mathbf{o}$ are zero by definition outside the region, i.e., $\Delta o_i = 0 \ \forall i \notin \mathcal{S}$. Thus, the gradient vector in Eq. 10 can be rewritten as $\nabla_{\mathbf{o}_\mathcal{S}} \chi_\mathcal{S} = \overline{\mathbf{W}}_\mathcal{S}^T \mathbf{p}_\mathcal{S}$ where $\overline{\mathbf{W}}_\mathcal{S}$ is the square sub-matrix of $\mathbf{W}_\mathcal{S}$ corresponding to the zones of $\mathcal{S}$. $\nabla_{\mathbf{o}_\mathcal{S}} \chi_\mathcal{S}$ is a vector of size $|\mathcal{S}|$ that denotes the marginal impact of a zone's change in opportunities to the overall regional accessibility. We refer to this vector as the 'opportunity improvement potential' of the region $\mathcal{S}$.

Notably, while this improvement potential is computed by using the regional population distribution, $\mathbf{p}_\mathcal{S}$, we posit that its computation can be modified. This modification would involve giving larger weights to socio-economically disadvantaged neighborhoods compared to more affluent areas. This aims to address equity concerns typically encountered in utility maximization approaches for solving the investment placement problem.

We demonstrate this effect for Cook County, the principal county of the Chicago MSA. Within this county, the region of southern Chicago has a long history of racial segregation, gentrification, low upward mobility, and heavy dependence on public transit [53]. It also suffers from high poverty, a high rate of single-parent households, and fewer high-earnings job opportunities (see Suppl. Figure 3). To locate the zones of this region, we compute a Socio-Economic Disadvantage Index (SEDI) of each BG as the regional quantile of measures of six socio-economic factors – poverty, racial minorities, unemployment, low education, vehicle ownership, and single-parent households. The computational details of this index are provided in Suppl. Section 1.6.

The distribution of the SEDI values across BGs of Cook County is shown in Figure 5a. We define the South Chicago region as the region shown in a thick black boundary based on visual inspection of the high SEDI zones in Figure 5a. SEDI is a rank measure, therefore the same zone can have different values depending on the region of choice (see Figure 5a1 for illustration). The modification of the regional population distribution, $\mathbf{p}_S$, is guided by the weights derived from the SEDI emphasizing areas with greater socio-economic challenges.

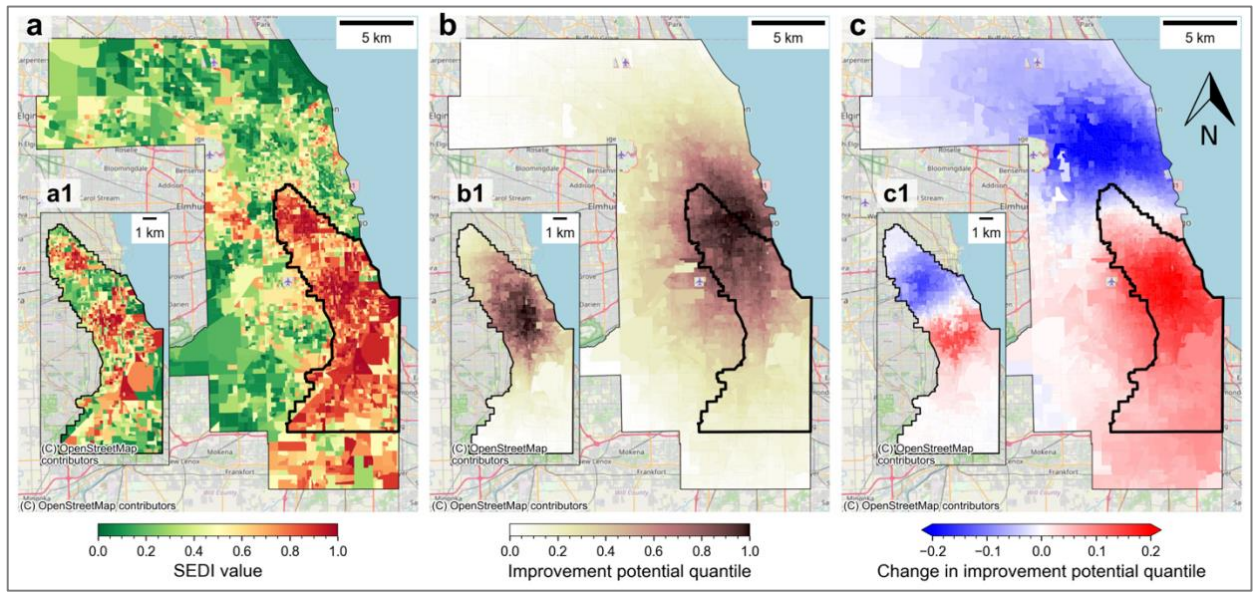

Figure 5: Relationship between socioeconomic disadvantage and opportunity improvement potential. (a) SEDI values at BG level in Cook County, IL, highlighting the selected South Chicago region; (b) Improvement potential of BGs in Cook County for 30-minute accessibility to essential shopping stores; (c) Difference in the improvement potential when measured with respect to unweighted population. The corresponding rank values computed separately for the South Chicago region are shown in the insets.

The spatial distribution of each BG's marginal impact on the overall accessibility of Cook County and the South Chicago region (in the inset) is shown in Figure 5b. This improvement potential corresponds to accessibility by driving within 30 minutes to POIs classified as 'essential shopping' stores (mainly grocery stores). Notably, despite applying higher weights to neighborhoods with high SEDI scores, the core areas near downtown Chicago show a higher improvement potential than the South Chicago neighborhoods. However, the prioritization of high SEDI block groups certainly leads to a difference in the ranks of the accessibility values. This is observed in Figure 5c which shows that southern Chicago is ranked much higher for investment opportunities compared to northern Chicago if the zonal populations are weighted by SEDI. Interestingly, areas close to downtown Chicago remain largely uninfluenced by this weighting of population. This could point to the dominant effect of current infrastructure and land use over any population weighting for equity purposes. These observations point to the profound influence of existing service disparities and land use patterns, which are substantially different from the distribution of communities along socio-economic lines. This pattern highlights the complex interplay between urban infrastructure, service availability, and socio-economic factors in shaping regional accessibility. This analysis serves as a tool for planners to visualize and prioritize investments in areas that would benefit most from an equity-focused approach.

## 5 Summary and discussion

Accessibility measurement is an important task in urban and transportation planning that lacks standardization and availability of granular data at large scales. In this study, we have developed a generalized locational accessibility metric building on past research that counts the number of opportunities of a specific kind feasibly accessible within a given time threshold, depending on the mode of travel and trip purpose. With the help of publicly available datasets, we have prepared accessibility data at a granular level for the United States and made it publicly accessible.

The proposed metric presents several advantages over conventional locational accessibility metrics. First, it combines the decay impedance functions of gravity measures with the idea of the cost threshold in the cumulative opportunities approach which makes it more realistic and intelligible. It is shown that using a constant impedance weight as used in contour measures can significantly overestimate accessibility compared to a probabilistic decaying impedance function. This overestimation is observed to be the highest in the densest areas of cities when measured in terms of total accessibility. However, this overestimation is the highest in the remote, sparsely populated areas of metropolitan areas when measured in relative terms. Since urban planning agencies operate within their jurisdictions and would thus be interested in relative measurement, the lowest drop in accessibility in the core urban areas would encourage planners to focus on land use compactness for accessibility maximization over improving travel speeds.

Second, this metric is generalized to include different types of destination kinds and trip purposes, including jobs (overall and segmented by income, industry, etc.) and non-workplaces, especially those considered essential, such as grocery stores, hospitals, and schools. We have shown that, unlike prevalent accessibility studies, it is important to consider different impedance functions for different trip purposes since different purposes warrant different importance given by travelers to the trip duration.

Third, the simple functional form of the proposed metric and its aggregation over larger areas lends it two useful properties – being able to measure transportation efficiency and consider equity effects. We introduce an accessibility-based measure of transportation efficiency and compare it between Houston, New York City, and Chicago metropolitan areas as a case study. The analysis shows that the high utilization of the transportation network in downtown areas reaffirms the importance of compact cities, while also showing a very high capacity of bicycle infrastructure to improve compared to walking and driving. Also, the current infrastructure seems to be primarily targeted towards commuting by driving, leaving a significant scope for improvement of essential service facilities. Secondly, the proposed metric makes it easy to assess the marginal contribution of the supply potential of the smaller zones or neighborhoods of a city or county to its overall accessibility. Planners can weight socioeconomically disadvantaged neighborhoods more to modify this marginal contribution and therefore help update investment decisions to prioritize overburdened and underserved communities.

Along with these advantages, we also recognize the limitations of this metric and the possible extensions that can greatly help standardize accessibility measurement in practice. First, it should be noted that it is a locational measure of accessibility, meaning it does not natively support including idiosyncratic travel preferences and situations. Second, the current form does not include competition effects which are commonly assumed to be true for facilities with much higher demand than supply. Further, data limitations prohibited us from computing this metric for public transit and variable times of day and/or days of the week. Finally, we hope further research provides more support on its ability to be integrated with existing transportation project economic analysis frameworks.

## Declarations

### Declaration of interest

This study was not sponsored by any funding agency. The authors do not declare any conflict of interest.

### Declaration of generative AI

### Acknowledgments

The authors would like to thank SafeGraph Inc. for providing the points-of-interest data through an invite during their public release of data during COVID-19.

**Code and data availability**

All analysis for this study is done in Python along with additional tools like `osmium-tool`. The code used to collect and process all the data is available on https://github.com/rvanxer/spr_4711. This repository includes a data file that includes the values of processed accessibility at the block group level in the contiguous United States by three modes, along with a brief description document.